\documentclass[pre,twocolumn,showpacs,preprintnumbers]{revtex4-1}

\usepackage{amsmath,amssymb,latexsym}

\usepackage{graphicx} 

\begin{document}

\title
{ Fertilizer usage  and  cadmium  in soils, crops and food.
}

\author
{M.W.C. Dharma-wardana\email[Email address:\ ]
{chandre.dharma-wardana@nrc-cnrc.gc.ca} }
\affiliation{
National Research Council of Canada, Ottawa, Canada, K1A 0R6
}
\affiliation{
D\'{e}partement de Physique, Universit\'{e}
 de Montr\'{e}al, Montr\'{e}al, Qu\'{e}bec, Canada.}

%
\date{\today}
%

\begin{abstract}
Phosphate fertilizers were first implicated by Schroeder
and Balassa in 1963 for  increasing  the Cd concentration in cultivated soils and crops.
This suggestion has become a part of the accepted paradigm on soil toxicity. Consequently,
stringent fertilizer control
programs to monitor Cd have been launched.  Attempts to link Cd toxicity
and fertilizers to chronic diseases, sometimes with good evidence, but mostly
on less certain data are frequent.
A re-assessment of this ``accepted'' paradigm is timely, given
 the larger body of data available today. The data show that
 both the input and output of Cd per hectare from fertilizers are negligibly small
  compared to the total amount of
 Cd/hectare usually  present in the soil itself. Calculations based on current
agricultural practices are used to show that it will take   centuries to double the
ambient soil-Cd level,  even after
neglecting leaching and other removal effects.
The concern of long -term agriculture should  be the depletion of available phosphate
fertilizers, rather than the contamination of the soil by trace metals.
This conclusion is confirmed by showing that the  claimed correlations between fertilizer
 input and Cd accumulation in crops are not robust. Alternative scenarios that
 explain the data are presented.
Thus soil acidulation on fertilizer loading, and the effect of Mg, Zn, and F ions
 contained in
fertilizers are considered using recent Cd$^{2+}$, Mg$^{2+}$  and F$^-$ ion-association theories.
The protective role of ions like Zn, Se, Fe, etc., is emphasized, and the question of
Cd toxicity in the presence of other ions is considered.
These  help to clarify  difficulties  in the standard point of view. This analysis
does {\it not modify} the accepted views on Cd contamination by airborne delivery, smoking,
and industrial activity, or  algal blooms caused by phosphates.
\end{abstract}

\maketitle

\section{Introduction}
\label{intro}

That fertilizers could be a serious source of Cd contamination of agricultural soils, and consequently
 the diet, was suggested almost half a century ago by Schroeder et al~\cite{Schroder63}. This view has now become a
 mainstream paradigm~\cite{McLaughlinSingh99,Loga2008} that has raised much public
 concern~\cite{EU-SoilContam2013,TothHeavyMet-MALs2016}, as also with the
 overuse of pesticides~\cite{Lechenet17Pesti}.
 The presence of Cd  in the environment, augmented by
 industrial activity, years of  coal and fossil-fuel usage,
 mining etc., is a serious health hazard and its monitoring  is essential, given its known  accumulation
 in the food chain with the potential for causing chronic diseases of the renal, pulmonary, cardiovascular
 and  musculoskeletal systems~\cite{ATSDR2013, JarupCd98}. 
 However, controversy exists regarding a number of aspects~\cite{CHANEY12,Roberts14}, and a
 re-assessment of
 the ``accepted'' view of Cd enrichment of soils by Cd in P-fertilizers is timely, given
 the larger body of data available on fertilizer use~\cite{Sillanpaa92,WorldBankFert}.  

Many tropical agricultural communities
 (e.g., in India, Sri Lanka, El Salvador, Nicaragua, Egypt, China) are facing a
 new type of chronic kidney disease of unknown aetiology (CKDu) 
appearing even though  recognized causes (e.g., diabetes,
hypertension, etc.) are absent~\cite{WHO2,Gifford2017CKDuReview}. Such CKDu is also found in
the developed world including Canada~\cite{Canada1}. Some authors have suggested 
Cd and other heavy metals to be factor causing 
such chronic kidney disease~\cite{Kamani12,WHO2,Mott-Abey13}. However,
 the  existence of CKDu communities adjacent to non-CKDu communities subject
to similar agrochemical exposure
is consistent with other explanations~\cite{DissaFluo05,IllepAl,SynergyBandara2017, CDW17Multiple}.
Traditional  agricultural communities have a relatively low fertizer usage.
For instance,  in 2002 El Salvador (which has CKDu) used about 71 kg/ha  while
New Zealand (no significant CKDu) used 1836 kg/ha according to~\cite{WorldBankFert} data.
 These show an
anti-correlation with fertilizer use and chronic disease, but many authors readily implicate
the ``green revolution'' and P-fertilizers
for chronic health issues of unknown aetiology, e.g.,~\cite{BandaraJM2010}.

In this study we deal mainly with Cd toxicity, while our discussion can be easily adapted to 
other heavy metal contaminants as well. We review the evidence and counter-evidence that exist to
 claim that increased fertilizer usage is correlated with increased metal-toxin levels in the soil,
 together with an increase of Cd  in  crops grown in such soil.
 International  regulatory bodies have  set a 60-70 $\mu$g tolerable maximum daily  intake
 for an average  adult~\cite{JECFA2011}. However, some societies  traditionally
 consume rice, or sea-food in diets  with Cd exceeding such limits, 
while  remaining quite healthy \cite{Sirot08Shell}.
 Hence, noting possible  counter-action among heavy-metal contaminants and micronutrients,
 a simple model for joint toxicity effects is considered in the last part  of this study.
 
Phosphates have been mined for the last 150 years. 
It is a non-renewable resource that must be mined from nature and  cannot be artificially produced.
This is a powerful argument for the reduction of mineral phosphate inputs in agriculture where possible. 
However, it is argued here that contrary to the commonly held paradigm that
 ``the addition of  phosphate
 fertilizers
 to the soil proportionately  increases the bio-available soil Cd'', simple mass conservation 
  limits any such increase to extremely small margins, well within the
 uncertainties of soil chemistry, bio-availability and  uptake of metal ions  by crops. It is suggested
 that
controlling the Cd content in fertilizers will have no discernible  effect on the
Cd content
in soil, and in crops. Hence the increasingly restrictive
efforts of some governments, esp. in the EU to minimize dietary  Cd
 inputs via fertilizer control
 will prove to be futile.  The European Food and safety Authority (EFSA) set the recommended tolerable
weekly level in the diet at 2.5 $\mu$g Cd/kg of body weight in 2012, and proposals to reduce the Cd
content in fertilizers correspondingly have been made. Even according to 2001 regulations
 more than a decade ago, the  amount of Cd allowed 
was  set at 400 mg/kg of fertilizer in USA (e.g., in Washington state, for, 45\% P$_2$O$_5$ product), while the
 EU countries proposed setting limits averaging close to  20 mg/kg of P$_2$O$_5$. In countries like Sri
 Lanka where public concern has been raised by groups pushing for `traditional agriculture', 
impractical limits as low as  3 mg/kg of fertilizer have been
 imposed with no basis in science~\cite{SLSI17}.
 Roberts, commenting on this restrictive trend remarks in 2012 that ``the rationale for the limits
  provided by the proposal provides little scientific evidence justifying a limit of 20 mg
 Cd/kg P$_2$O$_5$  and there is little evidence
 in the scientific literature suggesting that Cd would accumulate in soils through using P fertilizers
 containing less than 60 mg Cd/kg P$_2$O$_5$, much less pose human health risks''~\cite{Roberts14}.
 Similar views are found in recent risk-assessment studies by other authors, e.g.,~\cite{CHANEY12},
 or the Wageningen  University report~\cite{Rietra17Wegenin}. 

In the following we present further evidence
 against the conventional paradigm of Cd accumulation by fertilizer inputs, and examine mechanisms
 where  fertilizer addition into soils trigger {\it existing soil Cd} making it
bio-available to plants. Mechanisms like (i) the effect of increased acidulation due to fertilizer addition,
(ii) competition from ionic forms of  Zn, Se, Fe, (iii) salinity effects, 
(iv) dissolved organic carbon, soil and plant characteristic etc., are usually examined, but
in addition  we consider  ionic mechanisms due to added F and Mg ions, previously inadequately
treated  in discussions of Cd dynamics in soils. Hence, if the
analysis given here is found  to be valid on further investigation, Cd uptake by crops from soils
may also require controlling  the Mg,  and F content in fertilizers and in the soil and ensuring an  excess of bio-available zinc ions over bio-available Cd ions.

\section{Cadmium accumulation and fertilizer use}
\label{Cd-acum.sec}
Subsequent to the suggestion of  Schroeder and Balassa~\cite{Schroder63} that the use of phosphate
fertilizer leads to Cd accumulation in soils,  interest in monitoring soils for cadmium
grew rapidly, with  Kjellstrom reporting in 1979 that measured Cd levels in wheat doubled from
 1920 to 1979~\cite{Kjellstrom79Cd}, while Singh claimed in 1994 that the
application of phosphate fertilizer for a period of 36 years resulted in a 14-fold increase in
Cd content of surface soils~\cite{Singh94Cd}. A noteworthy step was the publication of the
soil bulletin No. 65 (FAO65) of the food and agriculture organization (FAO)~\cite{Sillanpaa92}, 
presenting
the status of Cd, Co, and Se in soils and plants of thirty countries, determined within
a uniform protocol enabling international comparisons. The publication by Silanp\"{a}\"{a} and Jansson (1992) sponsored  by the FAO
will be referred to  as FAO65 when convenient.  Data from Figure 5 given in FAO65
are shown in Fig.1,
where a clear correlation of the Cd content in the soil extracted
using a mild reagent (see below)  are shown for a time duration of three years of P-fertilizer
application. 
%
\begin{figure}[t]
\label{CdPlotA.fig}
\begin{center}
\includegraphics[width=0.95\columnwidth]{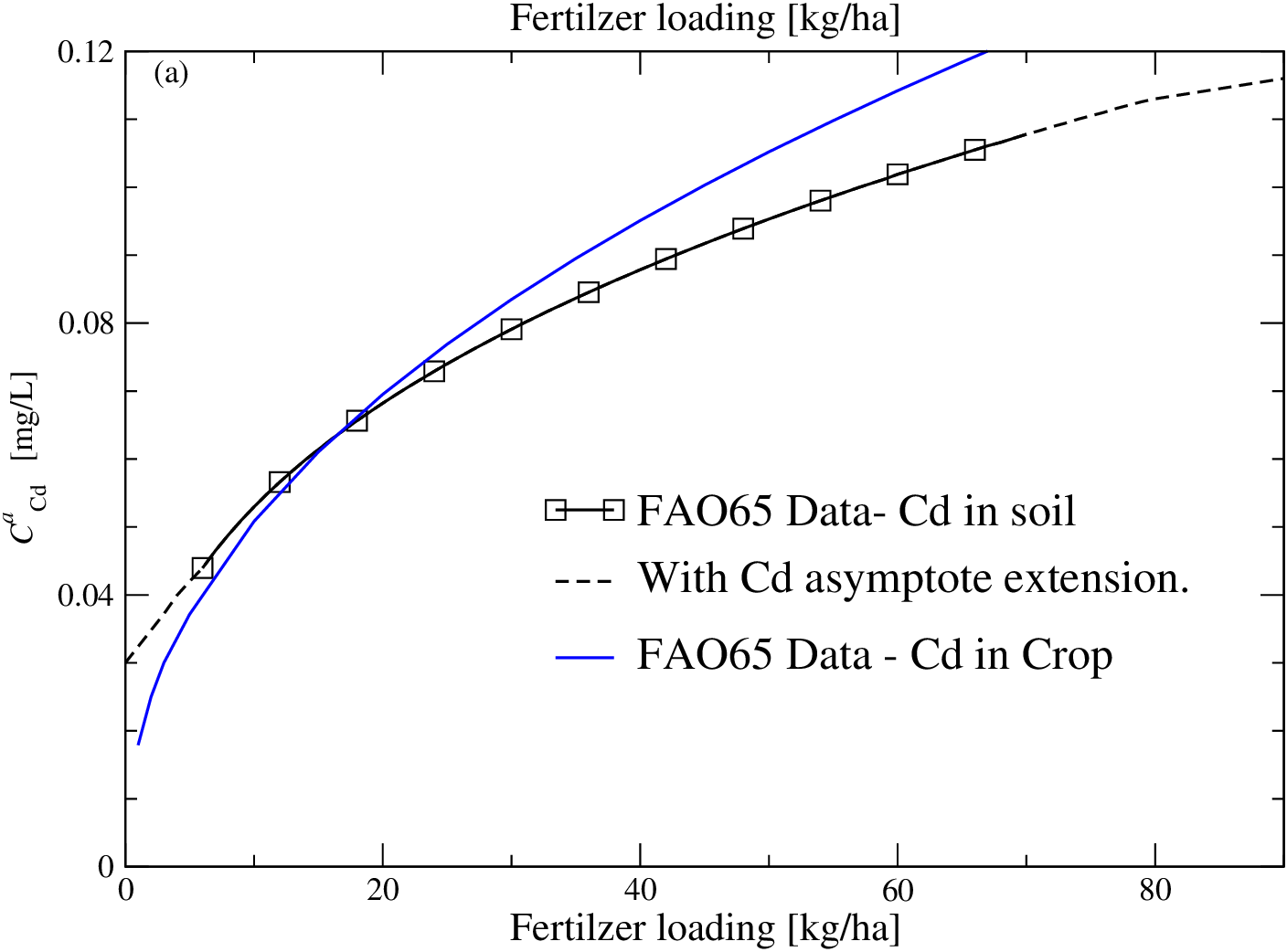}
\caption{(On line colour) Soil Cd (extracted using a mild regent) $C^a_{\rm Cd}$
 as functions of P-fertilizer
 input, (FAO Soil Bulletin No. 65, Sillanp\"{a}\"{a} and Jansson, 1992) and its extension
 (dashed lines)
using  asymptotic constraints.}
\end{center}
\end{figure}

Sillanp\"{a}\"{a} and Jansson concluded that ``although the fertilization data covers only three years,
 the relatively firm correlations leave no doubt as to the increasing effect of phosphorous fertilization
 on the  Cd contents of both plants and soils''. However, from this observation
it  also needs   one further tacit  {\it assumption} to conclude that this increased  Cd comes
 from the Cd impurities contained in the P-fertilizers {\it added} to the  soil annually. 
Sillanp\"{a}\"{a} and Jansson do not  in fact make that assumption, but many readers may easily do so.   
 This  may be called the ``Soil-Cd Enhancement by Fertilizers''  (SCdEF) assumption, and  is articulated
 quite clearly in many other  works  
~\cite{Mulla80Cd,McLaughlin96Review,Erik01Cd,Moolenaar99Budget,Jansson02,Premarathne2006,TothHeavyMet-MALs2016}. In the
following we argue that the data can be more correctly interpreted as a case of the P-fertilizer
triggering the conversion of already existing soil Cd to an  `available form' of Cd
 extractable by mildly acidic reagents. 

 Hence it is instructive to reexamine many of the studies of the period when 
the SCdEF paradigm was developed. 
McLaughlin {\it et al.} (1996) present a soil-balance calculation  in the caption to
 their Table 2 which states the following. ``Assumes 20 kg P/ha applied per wheat crop and 80 kg P/ha
 per potato crop and fertilizer contains (per kg P) 50 mg As, 300 mg Cd (250 mg Cd for potatoes),
 5 mg Hg, 200 mg Pb and 200 g F. Element inputs in irrigation water assumed to be negligible, although
 F may be a significant impurity in some waters''. They also assume a fertilizer-application depth
 of 10 cm
  of soil, taken to have a density of 1.3 kg per litre.  No leaching of the metal impurities
 added to the
 soil via the fertilizer is included, but such corrections can be easily applied.
 McLaughlin {\it et al.} (1996)  report analytical
 data for a variety of phosphate fertilizers. 
We have included a selection of these in Table~\ref{CdinFert.tab}  together with other data, e.g.,
 for Sri Lanka and India where some regions are affected by chronic kidney disease.
Columns 2 and three in the Table enable one to roughly convert among the various methods
 of indicating the Cd concentration
in rock-phosphates, viz., as mg/kg of rock, mg/kg of P, or mg per kg of  P$_2$O$_5$, with
the  P/P$_2$O$_5$ also dependent on the origin of the mineral although a factor of around 0.4 
is sometimes  used. In citing published work we have retained the  units used by the cited authors. 
\begin{table}
\caption{Cd and P concentrations in some rock phosphate sources for fertilizers.
The total P$_2$O$_5$ contents are very variable, 
ranging from 15-17\% (Russia, Chile) to 35-36\% (Senegal,
Togo), with an average of 27\%. Phosphate rocks containing less than 20\% P$_2$O$ 5$ need 
beneficiation  to justify transportation costs. A minimum of 25\% P$_2$O$_5$ is usually required.
}
\begin{tabular}{l l l l l }
\hline\\
Source                               & Cd        &     P     &  Cd         &\\ 
                                     &[mg/kg]    &    \%     & [ mg Cd     &\\  
                                     &           &           &  /kg P ]    &\\  
\hline\\
Russia$^a$                           & 0.2       &    17     &  1          &\\  
China(Yunan)$^a$                     & 5         &    14     &  35         &\\  
Sri Lanka:     \\
$\;\;\;\;$(Eppawala)\textdagger      & $<3$      &    14     &  34         &\\
$\;\;\;\;$(imported)$^b$\textdagger  & 2.3-46    &    7-20   &  325        &\\  
India (Mussoorie)$^c$\textdagger     & 8         &    12     &  62         &\\   
Egypt$^a$                            & 8-9       &    13     &  61-67      &\\
Morocco$^a$                          & 12-34     &  14-15    &  88-240     &\\
USA (N.C.)$^c$                       & 33        &    15     &  240        &\\
Nauru (NZ)$^a$                       & 100       &    15.6   &  641        &\\
\hline

$^a$ \cite{McLaughlin96Review}\\ $^b$ \cite{Zapata04FAO}\\
$^c$ \cite{VanKau97} &
\textdagger  Estimated.\\
\end{tabular}
\label{CdinFert.tab}
\end{table}

For single superphosphate (SSP) produced by reacting phosphate rock with sulphuric acid, 
 produced by acting on phosphate rock with phosphoric acid, 
most of the Cd in the phosphate
rock is transferred to the SSP. In wet-process phosphoric acid (WPA), about 55-90\% of the Cd is
transferred to the acid with the balance to the gypsum (a by product). Ammonium phosphates
(e.g. monoammonium phosphate [MAP] and diammonium phosphate [DAP]) are produced from WPA.
Their Cd content can range from $ < 1$ to $> 100$ mg/kg, depending on the mineral.

\subsection{\bf Cadmium input into the soil  on application of phosphate fertilizer.}
\label{CalcCd.sec}
In order to examine more closely the validity of the  SCdEF assumption, we recalculate
the incremental change in the soil-Cd concentration, $\Delta C^s_{\rm Cd}$ on addition of P-fertilizer
 to the soil.
We summarize the result using the symbols $A^F$ for the amount of fertilizer (kg/ha) applied annually,
$C^F_{\rm Cd}$ for the concentration of Cd (mg/kg) in the fertilizer, $d_s$ the depth
of the soil layer in cm., while $\rho_s$ is the density of the soil in kg/litre. The
total concentration of soil Cd is denoted by $C^s_{\rm Cd}$. Then the change
$\Delta C^s_{\rm Cd}$ on fertilizer loading is:
\begin{eqnarray}
\label{gperkg.eq}
\Delta C^s_{\rm Cd}&=& \frac{ A^F C_{\rm Cd}^F}{d_s\rho_s}\times 10^{-8}, \; \mbox{Cd, g/kg of soil}\\
\label{ngperkg.eq}
   &=& \frac{10 A^F C_{\rm Cd}^F}{d_s\rho_s}, \;\; \mbox{Cd ng/kg of soil}
\end{eqnarray}
The change of Cd concentration, being very small, is given in nanograms per kg of soil in Eq.~\ref{ngperkg.eq}.
We have ignored the additional inputs (e.g., via airborne Cd and via irrigation water) although airborne
Cd may be a major  source of Cd deposited on soils in industrialized countries. The Cd inputs via irrigation
water can be neglected in normal farming environments in most countries like the EU, Canada and USA, and 
even in less regulated non-industrial environments.

 For instance,  in a publication relating to CKDu in Sri
 Lanka~\cite{Dharma2015} the authors considered the non-point source
transport of phosphate by the irrigation waters of one of the major rivers (Mahaweli) of Sri Lanka,
but the amounts of Cd and other metal toxins transported in the same manner would be quite negligible,
being present in parts per million compared to macro-nutrients.
Thus,  consider one hectare
of the tea growing region where the rivers originate, with an average annual rainfall
 of 2 meters, and an annual average
fertilizer input of 100 kg/ha, containing 30 mg/kg of Cd (or any other such impurity).
 If half the rainfall
contributes to the river run off, the concentration of Cd from the fertilizer input
is only 0.3 parts per trillion. On the other hand, macro-nutrients
 (e.g., phosphates) are important pollutants that cause algae blooms.  
 \cite{Diyabalanage2016} confirmed  by  detailed analytical studies
 of Mahaweli river water that metal toxin levels are indeed below maximum allowed limits (MALs).
What is measured is the washoff from the {\em existing} soil cadmium, as
the contribution from fertilizer inputs is negligible.
 Similarly,  \cite{Jayasinghe-RO-15} showed that toxin levels in
irrigation waters  were well below the  usual MALs and hence required
no reverse-osmosis treatment to render them safe.
 \cite{McLaughlin96Review}  also disregard irrigation-water inputs
 of Cd into farm soils. A study of the translocation
and dispersion of pesticides by irrigation waters of the Mahaweli river also showed the effect
to be negligible~\cite{AravinnaPesti-17}. That this should be so is also easily ascertained
by a simple calculation based on mass balance. 
    
Essentially the same analysis as for Cd can be used for As, Pb and other heavy-metal additions to crops
 via fertilizers, be they wheat, barley, rice, or any other crop, and the concentration increment
$\Delta C_{\rm Cd}^s$ turns out to be in parts per trillion to fractions of parts per billion 
($\mu$g/kg of soil).
Only a fraction of this, would be bio-available. This is further lowered if we take into account
 any leaching effects of rain fall and irrigation wash-off (esp. under monsoonal conditions in
the tropical belt).

 Thus, even after a millennium of industrial
 agriculture using a typical rock-phosphate  fertilizer (see Table~\ref{CdinFert.tab}),
 the total Cd inputs remain negligible even for accumulations over centuries~\cite{CDW17Multiple}.

In contrast, the  calculations of the ``Cd budget'' given in  publications by various authors
usually extract a different conclusion that supports the SCdEF paradigm. The ambient total Cd in the soil, 
 $C_{\rm Cd}^s$ in European  soils (within the
`plough layer') can range from 0.05 mg/kg to  higher values (in industrialized areas e.g., 
in  Belgium, Hungary, see FAO65, i.e., Sillenp\"{a}\"{a} and Jensson 1992). Soil Cd amounts
 in Shipham, Wales, UK, ranges from 9 mg/kg
 -- 360 mg/kg \cite{EU-SoilContam2013}.
  A mean value of
 0.4 mg/kg is sometimes  used in model calculations for the EU~\cite{Rietra17Wegenin}, while 0.3 mg/kg
has been proposed by~\cite{Smolders13} as an average for the EU.
Scandinavian soils have a lower average of 0.2 mg/kg \cite{Erik97}. 

Interestingly, the Cd concentrations
 in the soil of  Sri Lanka are reported to range from 0.42 mg/kg in forest soils,
 to as high as 5 mg/kg in  lake sediments~\cite{Chandrajith12}, and are consistent
with values found in the WHO-sponsored study~\cite{WHO2}. 
However, most of the Cd in Sri Lankan soils is found as
bound Cd, since   the Cd contents in water and in soil solution were
found to be $< 3\mu$g/L and are below the MAL~\cite{WHO2}.

 ~\cite{TothHeavyMet-MALs2016} raises the interesting possibility that
the low values of Cd in E. Europe, in comparison to W. Europe, are  possibly due to the use of Russian
P-fertilizer in E. Europe, as opposed to Moroccan fertilizer used in W. Europe.
 However, the calculations presented in Sec.~\ref{CalcCd.sec}
 show that the Cd content of Moroccan P-fertilizer cannot account for such a difference.
The high content of soil Cd in industrialized regions (e.g., in W. Europe) should be attributed to
industrial activity, coal-power production,  and Cd
deposition from emissions. These are far more important than Cd inputs via P-fertilizer
applications.  Fortunately, according to ~\cite{Smolders13} airborne Cd sources have decreased
 by a factor of
five between 1980 and 2005. They propose a 0.35 g ha$^{-1}$y$^{-1}$ as the mean Cd airborne deposition
rate for the EU region currently.  We limit our study to soil Cd and Cd from
fertilizer inputs.

In order to understand the difference between our conclusions  and the traditional approach to the
soil budget for Cd, we review such a calculation~\cite{Erik01Cd} for southern Sweden
extracted from the Doctoral thesis of Jansson (2002), Table 1, column 2. 
 Eriksson considers the Cd inputs and outputs (g ha$^{-1}$y$^{-1}$) in his Cd budget.
\begin{enumerate}
\item {P-fertilizer, 0.12g from 10 kg P ha$^{-1}$ containing 12 mg Cd kg$^{-1}$} of fertilizer.
\item {Deposition: 0.7g  from airborne sources, rain etc. (Note that ~\cite{Smolders13} proposed
a 0.3 g annual addition from deposition as an EU average in 2013). }
\item {From lime, 0.02g added for soil remediation.} 
\item { Hence total Cd input = 0.84 g ha$^{-1}$y$^{-1}$.}
\end{enumerate}
Cd removal from soil is evaluated as follows:
\begin{enumerate}
\item {crops, 0.23g by plant uptake, removal of roughage, stubble etc.}
\item {leaching, 0.40g Cd, assuming a top soil layer  25 cm deep.(N.B., much
higher leaching rates are proposed in recent studies as European averages, e.g., in ~\cite{Smolders13}).}
\item {Total amount removed =  0.63g.}
\end{enumerate}
This leads to a total accumulation of 0.21 g ha$^{-1}$y$^{-1}$, 1/3 of which is due to
deposition. The amount that may be claimed for P-fertilizer is 0.12 g ha$^{-1}$y$^{-1}$,
and this is taken to support the SCdEF paradigm, leading to the conclusion that
accumulation of Cd impurities in fertilizers pose a serious health risk. 
However, this accumulation occurs in a soil volume 25 cm deep over an area of a hectare,
 i.e., in a soil volume of 25$\times10^5 $ liters, corresponding to
 a soil weight of 3.25 $\times 10^6$ kg with a soil density of 1.3 kg/L,
producing a {\it change} in Cd concentration $\Delta C_{\rm Cd}^s= 43\times 10^{-9}$ g/kg
of soil, i.e., a change of the order of 40 ng/kg which is truly negligible.
 Thus Eriksson's Cd budget, and those of other
workers  are  consistent with our  calculation giving mere nanogram/kg changes
in Cd {\it concentration} in the soil. The mean median Cd concentration in top
soils (0.2 mg/kg) and  subsoils (0.1 mg/kg) as reported  by  ~\cite{Erik97}
are trillion times bigger. The total soil Cd in the plough layer is 650 kg/ha.
Hence the parts per trillion increase in Cd concentration due to fertilizers is
negligible. Unlike the 0.12 g ha$^{-1}$y$^{-1}$  Cd  input of the P-fertilizer,
 the 0.7 g ha$^{-1}$y$^{-1}$  airborne deposition of airborne Cd
does not necessarily get ploughed into a 25 cm deep soil layer, but  affects
a few centimeters of the  topmost layer, causing more drastic changes in the
 soil-Cd concentration in the near surface.

%
%

McLaughlin et al (1996), and Loganathan et al (2008) have given estimates for
the doubling of the background soil Cd and  F due to fertilizer addition.
We take  ~\cite{Loga2008} as an example of such  calculations and
review their   table 2 with suitable complementary information
in  Table~\ref{LogaTab2.tab}. It is of course not the doubling of the ambient
levels that matter, but their reaching the maximum allowed limits (MALs) for
 agricultural
soils. However, since the determination of the MALs is itself an uncertain
toxicological issue, the doubling time is an important measure.

\begin{table*}[t]
\begin{center}
\caption{Estimated time for doubling the concentrations of Cd and fluoride as
given by Loganathan et al (2008), assuming a soil density 1.0 kg/l, a soil depth of 10 cm.,
and ignoring small corrections due to pasture \& animal uptake,  removal, leaching, etc. }
\label{LogaTab2.tab}
\begin{tabular}{l l l l l l l}
\hline\\
Element   & Input of    & Equiv. kg   & X in     &   rate of   & soil           &Years to \\
  X       &  P-Fert.    & of rock    & fert.     &   x added.  &  conc. of X    &double the  \\
          & [kg p/ha/y] & fertilizer.&  [mg/kg]  &  [g/ha/y]   & [mg/kg ]       & conc. [y] \\
\hline\\
Cd        & 30          &  200       &  42            & $\sim$8.4      & 0.3          & 36 \\
F         & 30          &  200       &  30,000        & $\sim$6000     & 300          & 51 \\
\hline
\end{tabular}
\end{center}
\end{table*}

We re-evaluate the above data using fertilizer inputs based on current farming practices where smaller
amounts of fertilizer are used, since the fertilizer input is  based on the soil Olsen-P levels
as well. The amount of  P-fertilizer/ha/y needed depends on the target harvest/ha as well as on the
available phosphorous in the soil (Table~\ref{harvest.tab}). Furthermore, since a large fraction of the
world consumes rice, we have included recommended fertilizer additions, as given by the Dept. of Agriculture,
Sri Lanka (DOA-SL, 2016), noting that TSP is recommended for paddy cultivation (3-4 month
irrigated crop) as the phosphate is needed in the short term.
Current agricultural practice is to use a mixture of mineral fertilizers and other 
(e.g., compost) fertilizers to get higher yields.  
Hence based on Table \ref{harvest.tab} we have chosen the amount of
fertilizer to be a phosphate fertlizer equivalent to 70-100  kg/ha rock phosphate,
 which corresponds to the 15-9 Olsen-P type of
soil for potatoes. Wheat needs about a fourth of the amount of fertilzer and hence
we may assume it to be similar to paddy cultivation discussed below.  In practice, the harvest
can be significantly increased by using a mineral+compost mixture, without going to
 higher mineral-fertilizer inputs. Similarly, targeted application of fertilizer to the root zone,
or banded-application methods can be used to reduce the needed fertilizer input,
as is increasingly the current agricultural practice.  

\begin{table}[b]
\begin{center}
\caption{(a) P-fertilizer requirement as a function of (Olsen-P) Phosphate availability in the soil
and the target harvest for potatoes, based on agricultural practices in southern Ontario and
northern Minnesota~\cite{MinFert18}. (b) Recommended TSP fertilizer for rice cultivation in
Sri Lanka as a function of Olsen-P phosphate (mg/kg) in the soil (DOA-SL, 2016). Note that
the P-fertilizer amounts for paddy  cover two rice growing seasons of the year.   
\label{harvest.tab}
}
\begin{tabular}{l l l l l l}
\hline\\
$\; \downarrow$  target harvest$^a$  &  Olsen-P$\to$  &  0-3 & 4-7 & 8-11 & 12-15  \\
30-35 tonnes/ha/y potato             &  [ppm]         &      &     &      &      \\
\hline\\
Fertilizer$^b$,  as P$_2$O$_5$       &[ kg/ha/y]$\to$   &  112 & 82  & 55   &  28     \\
             as P                    & $\;\; $  ,,    &  42  & 31  & 20   &  10     \\
             as Rock P               & $\;\; $  ,,    &  280 & 205 & 135  &  70   \\
\hline\\
Paddy cultivation                    & Olsen-P$\to$   & 0-5  & 5-10& 10-15& 15-20  \\  
\hline\\
Triple Super Phosphate               &[kg/ha/y]          & 70   & 40  & 0    & 0\\
\hline\\
\end{tabular}
$\,$\\
$^a$ Increased harvest, e.g, 45 tonnes/ha/y can be obtained by\\
 using the inputs recommended for the lower Olsen-P range.\\
$\;^b$  P$_2$O$_5$ concentration in concentrated superphosphates $\sim$ 45\%,\\
while typical rock phosphates may contain about 38-42\%. 
$\,$\\
\end{center}
\end{table}

Hence  we use the figure of 70-100 kg/ha/y of rock phosphate (RP) or 15 kg P/ha/y
 as our nominal fertilizer input  for potato  cultivation for typical
 soil Olsen-P levels.
Rock phosphate will be assumed to contain 30 mg/kg of Cd, and 30 g/kg of F, (e.g., as in Moroccan RP).
In the case of paddy cultivation, we take 40 kg/ha/y of TPS as the typical
input, based on DOA-SL (2016) specifications for both cultivation seasons.
 Since the conversion of RP to TSP leads
 to the transfer of a part of the Cd and F content to
gypsum and other byproducts, the Cd and F content in the TSP will be taken
 as 15-20  mg of Cd per kg and 15-20 g F
 per kg of TSP respectively while the parent rock phosphate may have contained 25-30 mg/kg.
 Hence our  calculations which revise those of Loganathan et al (2008)
are given in Table~\ref{CdF.tab}.
\begin{table*}[t]
\begin{center}
\caption{Estimated time for doubling the concentrations of Cd and fluoride 
using current farming inputs for potatoes, and for paddy, assuming a soil
 density 1.3 kg/l, a soil depth of 20 cm.,
and ignoring  corrections due to pasture \& animal uptake,  removal, leaching, etc.
In the case of paddy cultivation, leaching and run-off due to monsoonal rains can be
considerable and increases the time for doubling ambient concentrations.}
\label{CdF.tab}
\begin{tabular}{l l l l l l l}
\hline\\
Element   & Input of    & Equiv. kg   & X in     &   rate of   & soil           &Years to \\
  X       &  P-fert.    & of rock    & fert.     &   x added.  &  conc. of X    &double the  \\
Potato    & [kg P/ha/y] & Fertilizer.&  [mg/kg]  &  [g/ha/y]   & [mg/kg ]       &conc. [y] \\
\hline\\             
Cd        & 15          &  100       &  30            & 3.0      & 0.3          & 260 \\
F         & 15          &  100       &  30,000        & 3000     & 300          & 260 \\
\hline\\
Paddy & [kg/ha/y]&  -      &  [mg/kg]         &  [g/ha/y]        & [mg/kg]        & [y]  \\
\hline\\
Cd        & 40 (TSP)      &  -      &  20            & 0.8            & 0.3          & 975 \\
F         & 40 (TSP)      &  -      &  20,000        &  800           & 300          & 975 \\
\hline\\
\end{tabular}
\end{center}
\end{table*}

Currently, there is greater effort to reduce fertilizer usage than in an earlier era,
and   our calculations  (using quantities conforming to current
usage patterns) give room for greater optimism. It may take time scales of a millennium
 to double the concentration of  Cd in soils under paddy cultivation, even if we neglect
removal processes (leaching, monsoonal runoff,
 and removal when crops, roughage, straw etc., are taken away from farmland).  Even in the case
of potato farming, even if we adopt 200 kg/ha/y rock-phosphate inputs (as in Loganathan et al 2008),
it will take over a century to double ambient soil concentrations of Cd or F. When such time scales
 are considered, the effect of other ions and the incorporation of Cd, F into bio-unavailable
 forms in the clay need to be considered.  Hence the major concern of long-term agriculture
 should be the  depletion of stocks of P-fertilizer and not below-threshold contributions
 to the concentration  of trace metals  coming from fertilizer inputs.

Fertilzer inputs for wheat are a factor of 4 less than for Potato (McLaughlin et al, 1996).
 As seen from Table 4 and the associated discussion, fertilizer inputs for rice are  also
 low. Hence Cd and other fertilizer-based inputs can
be cut down by perhaps a factor of 4, there by increasing the soil-Cd doubling time to 
millenium time scales  by adopting the following steps.
\begin{enumerate}
\item{ Potato diets could be increasingly replaced by rice or wheat-based diets as the needed fertilizer inputs  
are much smaller.}
\item{ In additon, genetically modified potato cultivars which mimic rice-like fertilizer response
   may be possible.}
\end{enumerate}

Some caution must be used with published data. Page 27 of the  Wageningan
 study~\cite{Rietra17Wegenin}
states  that ``the average annual inputs of fertilizers to agricultural soils are of the
order of one to three g/ha/y. At a Cd level in soil of 0.4 mg/kg, assuming a rooting zone of
20 cm and bulk density of 1.2 kg/L, this amounts to a total Cd pool of approx. 960 g/ha".
In effect, the correct value is 960 kg/ha; we give this simple calculation
 for the convenience of the reader.\\

\begin{tabular}{lll}
1. Soil volume/ha  &=& 10,000 sq. m $\times$0.2 m, i.e., 2000 m$^3$.\\
                  &=& 2$\times 10^6$ litres.\\
2. Soil mass/ha   &=& (2$\times 10^6)\times(1.2)$ kg. \\ 
                  &=& $2.4\times 10^6$ kg.\\
3. Mass of Cd/ha  &=& $(2.4\times 10^6)\times(0.4\times10^{-3})$ kg.\\
                  &=& $0.96\times10^3$ kg = 960 kg.\\
\end{tabular}  

Thus the maximum 3 g/ha/y corresponds to a
a change  of about three parts per million, and not parts per thousand, as implied in the
Wageningen study.
Nevertheless the authors had correctly noted that ``reducing
the Cd load  by fertilizer would have a very minor effect on the Cd pool during the
first few decades ...''. In fact it can be further strengthened to say that there would be a very
minor effect even in centuries, rather than decades. That is, European data also give us room for
more optimism than the prognosis from  McLaughlin et al (1996), 
Loganathan et al (2008), and other pioneering studies.  A similar analysis  can be used to show
 that trace amounts of arsenic found in P-fertilizers have a negligible effect on the
 ambient concentration  of soil arsenic~\cite{CDW17Multiple}. Hence extreme public policies on Cd 
 and As content in P-fertilizers, driven by the SCdEF paradigm  cannot be justified by the
 available scientific data. In fact, the available world reserves of P-fertilizers would
probably run out long before the soils reach anywhere near the MALs for adverse  health effects.

Similarly, the conclusion by Jayasumana et al (2015) that trace amounts of Cd, As, etc., found in
agrochemicals have an environmental effect leading to chronic kidney disease is completely in error,
as the trace elements, when distributed in farm applications contribute mere fractions of parts
 per billion which are far below accepted MALs for chronic toxicity. 

Furthermore, the origin of the increased  Cd content detected in the soil, and in crops, c.f., Fig.1,
using a mild regent, cannot be due to the Cd coming from the fertilizer.
It is released from the soil itself, by the action of the fertilizer on the ambient soil Cd.
We examine this further in the next section.

\subsection{\bf The effect of  P-fertilizer on available soil Cd.}
\label{noeffect.sec}
The discussion in the previous paragraphs shows that modifying the Cd content in
the P-fertilizer, e.g., using a low-Cd fertilizer as opposed to a high-Cd fertilizer should
show no effect on the Cd levels available in soil solution to crops grown in most soils.
 In this section we give  experimental evidence
in support of this conclusion that we obtained from considerations of mass conservation. 

Figure 1,
usually invoked to support  the SCdEF paradigm
actually implies the opposite (see Sec.~\ref{pH-Cd.sec}). Soil Cd data reported in ~\cite{Sillanpaa92}
had been determined using the Acetate-Ammonium Acetate Na$_2$EDTA (AAAc-EDTA)
reagent which measures ``available'' or `easily extractable'  Cd rather than the total concentration
 of Cd per kg of soil. The plot shows that on application of 70 kg ha$^{-1}$y$^{-1}$ of P-fertilizer
for three years, the available soil Cd concentration had reached 0.12 mg/L, while a much
weaker loading at $\simeq 5$ mg/kg gives
 $C_{\rm Cd} \simeq $ 0.04 mg/L.  Taking the density of soil to be 1.3 kg/L, mean increment,
$\Delta C_{cd}^s=(0.12-0.04)/(3\times 1.3)$=0.0205 mg/(kg.year).
However typical $\Delta C_{cd}^s$ are in the nano to micro-gram range (at the most),
as seen from calculations given in sec.~\ref{CalcCd.sec} and  Eq.~\ref{gperkg.eq}.
  Hence this result is 20,000 to 20 times too
 large for it to have originated  from the Cd amounts that were
input via the  P-fertilizer used. As airborne Cd and other inputs were
excluded in the experiments reported in FAO65 (Sillanp\"{a}\"{a} and Jansson 1992),
 conservation of mass implies that
 almost all of it originated  from {\it the  pre-existing  Cd pool} in the soil,
 but initially not extractable using AAAc-EDTA. It is converted to  ``available'' Cd
by  some  mechanism (see below) activated by the agrochemical inputs,
and converted to ionic Cd accessible with the mild reagents like AAAc-EDTA, and by plants.
Of course, the total Cd concentration $C_{\rm Cd}^s$ can be determined by standard methods
using extraction with strong acids (e.g., 2M nitric acid). Experimental data
fitted to algebraic formulae  connecting  the AAAc-EDTA extractable  Cd,
taken to be of the form $C^a_{\rm Cd}= a+bC^s_{\rm Cd}$  or expressed as  log-scaled
regressions have  been quoted by many authors~\cite{McLaughlin96Review,Smolders13},
and in FAO65.

Field experiments showing that the Cd content of the fertilizer may have
little or no impact on the soil Cd concentration and on the crop-Cd
content are   found even in the literature of the period, and are  alluded to
 in reviews  by various authors, e.g., ~\cite{McLaughlin96Review, Grant02Cd}.
Here we  refer to  a number of such examples, \\
(a) Sparrow et al (1992)  compared
Cd uptake by potatoes fertilized with both low- and high-Cd DAP
 in field trials. They found little differences in Cd uptake between the two cases, with Cd
concentration in tubers being related to the rate of P applied, rather than to the
amount of Cd applied.

The Cd content of durum wheat fertilized with MAP containing
varying amounts of Cd (0.2, 7.8, and 186 mg/kg), and  grown at 11 different locations over
a three-year period had no significant dependence on the Cd content in MAP~\cite{Grant02Cd}.
Thus, in spite of soil differences in the 11 locations, the result remained robust.\\
(b) P-fertilizers are usually applied to the soil together with N and K
fertilizers.
 Nitrogen-fertilizers (e.g., urea, ammonium salts) are converted by soil bacteria
to nitrates, generating acids (H$^+$ ions)  in the soil, while base ions are transferred
to plants. Hence additional availability of Cd in the presence of phosphates may also
be caused by accompanying N fertilizers. This may be viewed as due to increased leaching of
 Cd fixed in the soil and conversion to Cd$^{2+}$ in soil solution, caused by decreased pH.

Already in 1976  Williams and David (as reviewed in McLaughlin et al, 1996)
showed that the concentration of Cd in  wheat grains harvested  from soils treated with
 superphosphate and ammonium nitrate  exceeded that with superphosphate  alone by a factor
 of two.\\  
(c) McLaughlin has also discussed the work of Sparrow et al (1992), and those of
 Williams and David (1977) where it is shown that the addition of P to a soil
 (with no change in Cd input) increases Cd uptake through
 a stimulation of root proliferation in the zone into which P is added.\\ 
(d) ~\cite{Onyatta05} report an increase of available Cd in the soil,
 induced by the use of P-fertilizer used in the form of Idaho mono-ammonium phosphate.
 They attribute their observations
to the release of pre-existing Cd from the soil as well as to the Cd coming from the fertilizer.\\
(e) Comparisons between Cd content in crops grown using commercial P-fertilizer, and organic
 fertilizers in field experiments for rye, carrots, potatoes showed no significant differences
in Cd levels~\cite{Jorhem00},

\section{Effect of  phosphate fertilizer on the soil.}
\label{Acid-soil.sec}
Soil is a complex subsystem containing clay, sand, organic materials loosely called `humus',
water, electrolytes, and dissolved gases, interacting with an interpenetrating  subsystem 
consisting of living organisms made up  of  micro-organisms,
insects, `bugs' and plants. The plants as well as the soil organisms need the mineral nutrients,
water as well as some of the organic matter for their existence, and exchange material
among them mostly via the soil. The exchange of nutrients between the  plant subsystem
 and the soil can be described
by transfer coefficients, and they need to be determined by experiment. 

Even the processes that occur entirely in the soil, e.g., the behaviour of the added fertilizer, and its
partitioning among clay, humus and the aqueous phase of the soil (called the `soil solution')
are  too complex for us to treat using first-principles atomistic models. Hence it has become
the practice to characterize the soil using various  macro-parameters of the soil, e.g.,\\
(a) the pH, soil texture characterized by a texture index (TI, see FAO65),
 organic matter content (OMC) $\gamma$ , cation-exchange capacity (CEC) $\xi$, salinity $\zeta$,
hardness $\eta$, and electrical conductivity $\sigma$;\\ 
(b) the concentrations of specific ions (micronutrients) like B, Cu, Mn, Mo, Zn etc.;\\
(c) elements toxic to humans, like  Cd, Pb, Hg, As.\\
However, while concentrations of micronutrients and  toxins are specified in `defining' a
soil,  macronutrients like N, K, P are not
specified as they are overwhelmingly controlled by fertilizer loading. Fertilizers themselves affect
the pH of the soil, and  nitrogen fertilizers  trigger soil-microbial action generating acids.
Hence the crop soils need pH adjustments which are usually achieved by the addition of 
 ag-limes like calcite and dolomite. 
\begin{figure}[b]
\label{ion-pair.fig}
\begin{center}
\includegraphics[width=0.95\columnwidth]{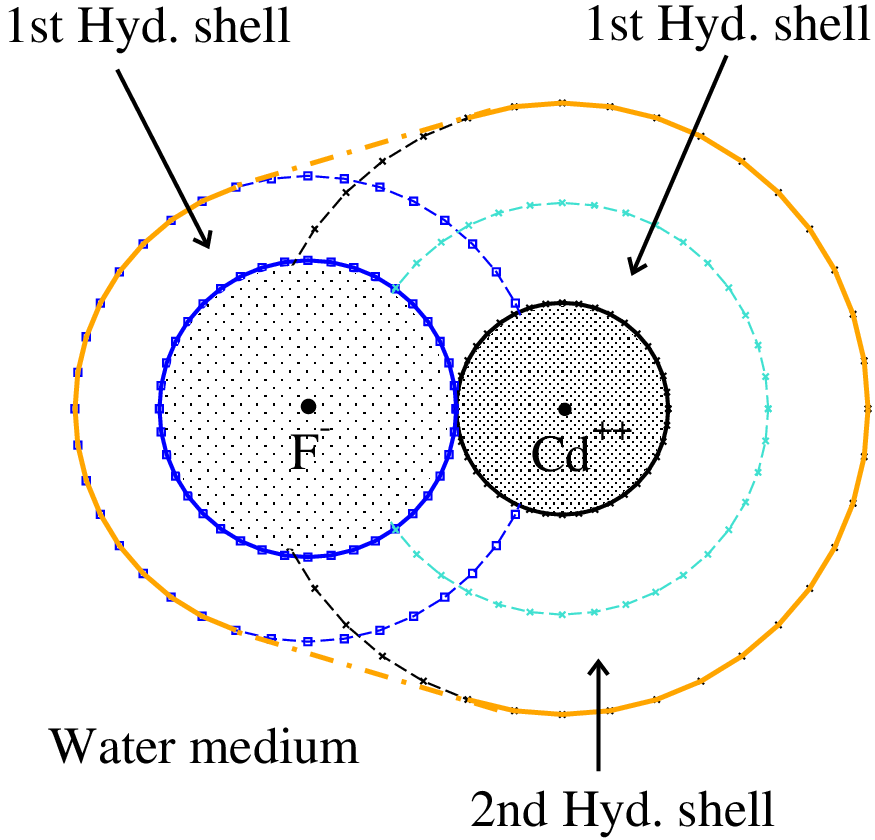}
\caption{(On line colour) A schematic diagram of the (Cd-F)$^+$ ion-pair together with the
 hydration shells of the  Cd$^{2+}$ ion and the F$^-$ ion prior to pairing.
The divalent Cd$^{2+}$ ion holds tightly two hydration shells, while the
`monovalent F$^-$ holds only a single hydration shell. The water outside
the joint hydration shell of the  pair `sees' an ion of effective charge
 $Z_p=1$. The reduction in solvation energy  on pairing
is offset by the paring energy when persistent ion pairs are formed.}
\end{center}
\end{figure}
An element like Cd can exist in several forms in the soil:\\
 (i) Cd ions chemically replacing Al or Mg atoms in octagonal environments, or
 replacing Si atoms in tetrahedral environments in clays. Here we are not
 implying any Cd-based crystal types, but only point-defect substitutions.
 These are soil-bound Cd with a concentration $C_{\rm Cd}^b$. They can be dislodged using strong
 reagents.\\
(ii) Fully or partially hydrated exchangeable  Cd ions electrostatically attached to edges,
 oxide groups etc., with a concentration $C_{\rm Cd}^{\rm xb}$. These ions may migrate into internal
sites with time, becoming strongly bound.\\
(iii) Fully hydrated Cd ions `available' in soil solution at a concentration
of  $C_{\rm Cd}^a$. These aqueous Cd ions
  carry a solvation sheath of water molecules. However stable associations
with other ions like fluoride forming a strong (Cd-F)$^+$ complex ion (see Fig.2) 
 can occur. 
It can be shown that such complexes are more stable than the hydrated Cd$^{2+}$ ion or
 the hydrated F$^-$ ion
 existing without ion association~\cite{CDW17Multiple}. In such cases, essentially all the
aqueous  Cd ions are in associated from since the F$^-$ concentration is largely in excess
 of the Cd$^{2+}$  concentration in most soil solutions. To complicate matters, such associated ions
 can also attach  electrostatically to edges, and surfaces of soil particles, and hence also contribute
 to $C_{\rm Cd}^{\rm xb}$.
 As they have a lowered positive
 charge, they are more weakly bound electrostatically. \\ 
 
Thus, while even the specification  of the concentration of Cd is complex due to the several
forms, viz. (i)-(iii), reagent chemistry can usually distinguish only between ``total Cd'' 
concentration
(extracted using strong acids), and the `available' Cd concentration extracted using milder reagents
like AAAc-EDTA. Their dependence on macro-soil parameters is obtained from field trials.
The use of such macro- parameters without using a  more microscopic physico-chemical
model of the soil implies that experimental data connecting them have to be linked by purely numerical
regression relations (curve fitting) containing coefficients without a clear  physical meaning. 
For instance,
the  bio-available or `accessible'  Cd concentration $C_{\rm Cd}^a$  in a soil
 measured (with a mild reagent)
as a function of the P-fertilizer loading $A^F$ can be  fitted to a
regression relation as given in Figure 5 of FAO65 (reproduced here in Fig.1 ),
\begin{equation}
\label{cd-fert.eq}
\log(C_{\rm Cd}^a) =-1.641+0.365\log(A^F).
\end{equation} 
Logarithms to the base 10 are implied.
Similar empirical relationships have been  constructed connecting other pairs of parameters like 
pH, OMC etc., but it is hard to assign error bars and domains of validity to them. Usually,
additional field trials fail to reproduce such fits in actual farm situations as additional 
factors weigh in. Furthermore,
the use of log-scaled parameters drowns much sensitivity, and renders such equations to be full of
pit falls if one were to use several equations in succession to eliminate variables and link
a pair of parameters which have not been directly fitted to experiments from field trials.
Nevertheless, currently used computer codes make wide use of such empirical fits and results of
``regression trees'' to provide  data bases  for algorithms whose outputs are rarely physically
 transparent. 

Another approach useful in colloid chemistry is to exploit
surface complexation modeling of  titration data on clean minerals like gibbsite, kaolinite,
providing rate constants for Cd absorption, retention etc~\cite{Weerasooriya02}. However, most such
experiments deal with Cd$^{2+}$ solutions in the 0.01 Molar solution range or higher, where as the ambient
exchangeable Cd levels in soil solutions are in the milli-molar regime (the bound
part of the  Cd pool may be 10 times larger in more alkaline soils). 
Nevertheless, as valid microscopic models are not available, we follow a strategy where
 empirical regression fits are judiciously used by constraining them to known asymptotic
 behaviour within simplified physico-chemical models.

It is instructive to look at a possible first-principles model of soil even though we will
not exploit it fully in this study. The clay  component can be modeled using
a crystal structure where tetrahedral SiO$_2$ sheets and octahedral sheets
(mainly Al or Mg oxide sheets with various cations replacing them) are the building blocks,
as in  montmorillonite (MMT), illite, or vermiculite. 
 The highly-reactive edge sites and surface defects control the stabilization of soil organic
matter, colloidal and rheological properties~\cite{Sposito08,Tombacz04}
The edges of the sheet structure of MMT-type clays represent the
boundary that solutes must cross in going between interlayer nanopores and
micropores. The dissolution of clay nanoparticles has been observed to proceed predominantly
from such  edge surfaces~\cite{Bickmore01}. Hence we may consider such  structures where Cd, Mg, Zn
and other ions may replace the Al ions in the MMT-type  octahedral sites, while some cationic
 substitutions of the tetrahedral Si sites are also possible. Hydrated ions can remain in
 the channels between layers, and constitute electrostatically held exchangeable
 cations in equilibrium with the cations  in the soil solution.

The addition of P-fertilizers and other agrochemicals can influence the ambient pool of Cd in
the soil in a variety of ways. These are:\\
(i) change of soil pH due to P-fertilizer loading, releasing soil-bound Cd into the soil solution, \\
(ii) change of concentration of competing ions like Zn, Ca, and micronutrient ions,\\
(iii) we examine the  effect of fluoride and magnesium added to the soil via fertilizer loading,
    and their ion-pair formation,
    topics which have not been adequately addressed in the past.\\
(iv)The effect of agrochemicals via their  ionicity and ionic strength in modifying the OMC of the soil.
    Here we use `ionic strength' as used in the theory of strong electrolytes, while 'ionicity' is
    used to indicate the capacity of an ionic mixture for denaturing or breaking up complex
    structures (e.g., in humus, proteins etc.) by the Hofmeister mechanism~\cite{Dharma2015, Baldwin96}.   

In the following we examine the first three items in greater detail.
 
\subsection{\bf Change of soil pH due to P-fertilizer loading, releasing Cd into the soil solution.}
\label{pH-Cd.sec}
Although the data given in the  FAO soil bulletin No. 65 (FAO65) are somewhat dated,
 they form a consistent set of continued interest for theoretical modeling.  Here we examine the
 data  given by ~\cite{Sillanpaa92}, reproduced here as Fig.2(a),
 using the regression equations
given  there to clarify possible mechanisms for the  increase in Cd content in the soil as
P-fertilizer is loaded over a time period.
\begin{figure}[]
\label{CdPlot.fig}
\begin{center}
\includegraphics[width=0.95\columnwidth]{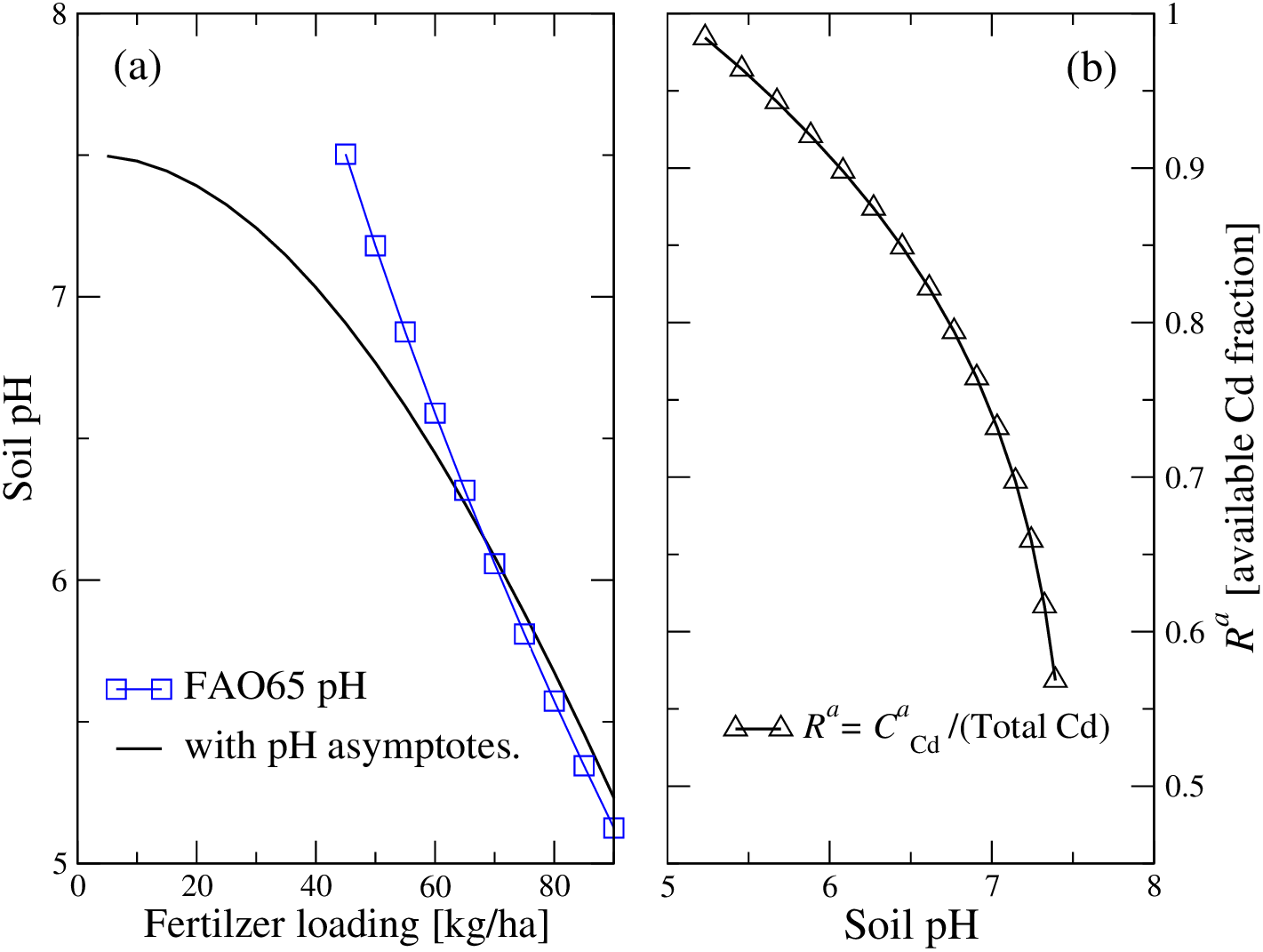}
\caption{(On line colour)
(a) Soil pH calculated using a regression relation between pH and $C^a_{\rm Cd}$ from FAO65 (Sillanp\"{a}\"{a}
and Jansson, 1992)and after
 constraining the alkaline regime (pH $>7$) to ambient natural $C^b_{\rm Cd}$. (b) The fraction of available
Cd in the soil, where the mean total soil Cd is taken as 0.12 mg/L of soil.}
\end{center}
\end{figure}
The amount of Cd in the soil available to plants depends on the
soil pH and its cation-exchange capacity (CEC), as the H$^+$ ions compete with the Cd ions
for  electrostatic binding to edges and surfaces of octahedral and tetrahedral building blocks of clays.
Also, decrease of pH   hydrolyzes  ions bound  to humic acids as they are weak organic acids.  
The data for the Cd content in soil used in Fig. 1, obtained  from ~\cite{Sillanpaa92} are
for Cd determined using the AAAc-EDTA reagent as described in
FAO65. This in effect extracts essentially the bio-available Cd,  with a concentration
$C_{\rm Cd}^a$, while strongly bound Cd located on clay sites are not extracted. Plants
are also able to serve themselves of this `available' Cd.
 Sillanp\"{a}\"{a} and Jansson (1992)  give the regression relation, Eq.~\ref{cd-fert.eq},
 connecting $C_{\rm Cd}^a$ with the the P-fertilizer loading $A^F$. Here we examine the extent of
of pH increase that is needed to explain these data (Fig.1) and if such a
pH-based  model is plausible. The rise in the pH associated with the loading of fertilizer
may be due to its intrinsic acidity (Table~\ref{CaMgFert.tab}), or due to bacterial actions
 triggered by increased availability of fertilizer.

\begin{table}[t]
\caption{Initial pH effect of some common fertilizers on soil acidity, 
and their nominal  Ca and Mg
content.}
\begin{tabular}{p{1.5 cm}p{2.5 cm}p{1.0 cm}p{1.0 cm} }
\hline\\
Source        & initial                   & Ca    & Mg   \\
              & pH-effect                 & \%    & \%   \\
\hline\\
MAP           & decreases to $\sim3.5$ & --    & --     \\
DAP           & increases  to $\sim8.5$   & --    & --     \\
Rock-P        & needs low pH to act       & 5-25  & 5-20  \\
SSP           & negligible effect         & 5-20  & 5-15  \\
TSP           &   ,,                      & 5-13  & 5-8   \\
Dolomite      & increases  pH             & 22    & 12    \\
\hline
\end{tabular}

Sources:~\cite{Uhawaii18} and Manufacturers' data sheets.
\label{CaMgFert.tab}
\end{table}

At the high-loading end we may assume that almost all the exchangeable Cd in the soil has
been released. The large-x asymptote to the curve, being a log-log regression is somewhat poorly defined,
as the fitting has not used such a boundary condition. Nevertheless, we can judiciously take it to
be close to the value attained at the highest loading, viz. $C_{\rm Cd}^a$=0.12 mg/L. Furthermore
FAO65 provides a regression equation connecting the available Cd and the pH, viz.,  
\begin{equation}
\label{cd-pH.eq}
C_{\rm Cd}^a=0.175-0.0111\mbox{pH}
\end{equation}
The pH used in this equation is the pH determined using a CaCl$_2$ buffer as defined in FAO65.
The use of this equation with Eq.~\ref{cd-fert.eq} by eliminating $C_{\rm Cd}^a$ is justified
only for pH$\ge 7$ as the AAAc-EDTA extraction becomes ineffective in alkaline media.  
At $C_{\rm Cd}^a$=0.10 mg/L, this equation predicts a pH of 6.1. Thus the acidulation needed
to achieve the observed increase in available soil Cd is eminently reasonable as
continuously  fertilized soils are known to reach  even higher pH (close to 4) unless remedied with
ag-lime. In order to model the higher (alkaline) range of pH, we assume (using the data in the
Appendix 1, FAO65)  that the unfertilized  initial soil had been adjusted to a pH of 7.5  at a zero
 fertilizer loading, viz., $A^F=0$, while the available Cd in the initial neutral soil
 is  25\% of the total available soil Cd.  In effect, we constrain the regression to
satisfy the (asymtotic) boundary conditions for  small x and large x. The resulting acidulation curve, i.e.,
pH due to fertilizer loading $A^F$ is shown in Fig.2(a).
\begin{table}[b]
\caption{Cadmium, selenium, zinc and fluoride concentrations in some rock phosphate sources
 for fertilizers. The indicated Zn concentrations are a lower bound.}
\begin{tabular}{p{2.5 cm}p{1.0 cm}p{1.0 cm}p{1.0 cm}p{1.0 cm} }
\hline\\
Source                 & Cd        & Se    & Zn     &  F      \\
                       &[mg/kg]    &[mg/kg]&[mg/kg] & [ g/kg] \\
\hline\\
USSR$^{a,d}$           & 0.1-0.2   &  n.a  & 19     &  n.a.   \\
Tunisia$^{a,d}$        & 38-53     &  11   & 385    &  41     \\
Morocco$^{a,d}$        & 3-34      &  3    & 209    &  n.a.   \\
USA (N.C.)$^{c,d}$     & 39        &  5    & 333    &  35     \\
Nauru (NZ)$^{a,d}$     & 100       &  n.a. & 1000?  &  30     \\
\hline
\end{tabular}

$^a$~\cite{McLaughlin96Review}; $^b$~\cite{Zapata04FAO};\\
$^c$~\cite{VanKau97};
$^d$~\cite{Bech10trace}
\label{ZnFinFert.tab}
\end{table}

The purpose of the exercise is to demonstrate that while we may qualitatively state that increased
acidulation of the soil triggered by fertilizer inputs can explain soil-Cd enhancement data like those
of the FAO65 set, they can in fact be
 addressed {\it quantitatively}  and the results are indeed quite plausible. However, while this
 might constitute
an explanation, it is by no means the only possible scenario that could lead to the observation
that the addition of  P-fertilizer to the soil increases the Cd available in the soil and hence
in crops grown therein. In fact, given that there are many factors affecting the concentrations
of available $C^a_{\rm Cd}$ and bound soil Cd $C^b_{\rm Cd}$ soil have to be given as a
function of al least the major variables. For instance, a popular empirical model is to use the form
\begin{eqnarray}
\label{Kd1.eq}
C^s_{\rm Cd}&=&C^b_{\rm Cd}+C^a_{\rm Cd},\;\; C^b_{\rm Cd}=K_DC^a_{\rm Cd} \\
\label{Kd2.eq}
\log K_D &=&a_1 + a_2 \mbox{pH}+ a_3\log C^b_{\rm Cd}+ a_4\log\gamma \\
         & &+ a_5 \log(C_{Clay}) + a_5\xi +\cdots,\nonumber
\end{eqnarray}

where five variables are included via the coefficients $a_i, i=1,2,\cdots,$. Equation \ref{Kd1.eq}
is written in the form of a mass-action law using the constant $K_D$ although this may not be
 justifiable as full equilibrium is
rarely attained. $C^b_{\rm Cd}$ consists of lattice-Cd atoms which may be embedded
in the tetrahedral -Si and octahedral -Al lattice sites of the clay particles, as well as Cd adsorbed to
 edges and surfaces of the nanopores and channels of the clay particles and humic acids. The 
adsorbed Cd  is likely to be in equilibrium with the ``available'' Cd present in the soil solution,
but not with the lattice-embedded Cd. In fact, no microscopic
model will lead to such a form as the above equation. In fact, Eqs.~\ref{Kd1.eq}, \ref{Kd2.eq}
 are really a testimony to our lack
of a quantitative understanding of the processes involved. When data are analyzed
 using such fits, in most cases one finds that the pH dependance associated with the fit parameter
 $a_2$ carries the dominant effect, providing a basis for the use of the simpler form given
 in Eq.~\ref{cd-pH.eq}.
  
As a counter argument to focusing on pH, we note that there are many inconsistent results
obtained in attempts to control the available soil Cd by  soil-pH remediation using, say,
ag-lime addition (e.g., see ~\cite{Jansson02} and references therein). 

\subsection{\bf Effect of competing ions like Zn, Mg, Fe, etc., on the available Cd in the soil.}
\label{zn.sec}
Equation \ref{Kd2.eq} does not make a serious attempt to take account of the effect of  other
ions like Zn, Mg, Fe etc on the Cd balance in the soil solution. Zn is in the same group of elements
as Cd in the periodic table, and has very similar chemical properties, with Zn being by far the more
reactive of the two. The radii of the hydrated Cd$^{2+}$, Mg$^{2+}$  and Zn$^{2+}$  ions are nearly equal,
 being about 4.2-4.4 \AA$\,$  depending on the aqueous environment. The evidence for such competition between 
Zn and Cd is widely available in the literature. In addition to their co-action in the aqueous `soil solution'
phase, they also compete for sites for incorporation in the inner substitution sites in octahedral
and tetrahedral locations of clay crystals. The ionic radii in the crystal lattice largely
 favours Mg (0.86\AA ),
then Zn (0.88\AA ) and least of all Cd (1.09\AA ). Thus long-term fixation by incorporation into
 the clay lattice applies for Mg, and Zn, but less so for Cd. More attention has been paid in the literature
to exchange with Ca$^{2+}$ ions
(radius in crystals, 1.14\AA), but its ionic radius is less favourable than
that of  Mg$^{2+}$ which is likely to have a larger impact on Cd dynamics
in the soil, as further discussed below.  


Nevertheless, effects of such competing ions are all lumped into the exchangeable cation term
 $\xi$ in Eq.~\ref{Kd2.eq}, and  in many Cd `risk-assessment' simulations. 
This  shortcoming is also reflected in the reports  of experiments on Cd in soil and in crops that
fail to report the amount of Zn present together with Cd, leading to inconsistent conclusions.
Greenhouse pot experiments using ``simulated'' fertilizer mixtures using pure phosphates and Cd
salts cannot be used to derive conclusions about actual farming outcomes where fertilizers typically
have a Cd/Zn ratio (Table. \ref{ZnFinFert.tab}) that may range from  1/10 to 1/100 ~\cite{CHANEY12}.
That is, Zn  largely dominates the Cd input from fertilizers and this effect cannot be ignored, or
lumped into a global `cation-exchange' term.

The role of Zn  has two contradictory effects. We examine them below:\\
 (i) In Sec.~\ref{CalcCd.sec} we showed that
the effect of the Cd input from P-fertilizers  can be neglected, but the Zn input, being possibly a
ten to hundred  times lager
than the Cd input, cannot be neglected, and has a strong impact on the {\it pre-existing} available soil
Cd $C^a_{\rm Cd}$ as the Zn-ions will free up many Cd ions (denoted by Cd$^{\rm bx}$) bound on
 to soil particles and humic acids moieties.
\begin{equation}
\mbox{Cd}^{\rm bx} +  \mbox{Zn}^{2+} \rightleftharpoons \mbox{Cd}^{2+} +\mbox{Zn}^{\rm bx} 
\end{equation}
The above equations must be coupled with the equation for the solubility product for the Cd$^{2+}$ and
PO$_4^{3-}$ equilibrium since Cd phosphate is relatively insoluble and the phosphate concentrations
in the plough layer are quite high, thereby suppressing Cd dissolution into the soil solution.

By making the assumption that the exchangeably bound zinc, Zn$^{\rm xb}$, and also the available
zinc (i.e, Zn$^{2+}$) concentrations are quite large compared to the corresponding Cd amounts,
the observed enhancement of available Cd on fertilizer addition displayed in Fig.1 can be explained
using a rate constant $K_D$ used in Eq. ~\ref{Kd1.eq}, with $K_D$ in the range of 1-100 depending on
 various reasonable
assumptions that one may make regarding the initial amounts of bound and available Cd, Zn etc., in the
soil prior to fertilizer application. Here we  keep the pH fixed as we wish to see if the data of
Fig.~1 can be explained purely in terms of the impact of Cd dynamics in the soil. As reported
 by ~\cite{Smolders13},  values for $K_d$ obtained by fitting to data bases can vary up to even 2300.
 Hence we see that the increase in available Cd concentration in soils as observed on fertilizer
 loading can also be accounted for quite easily by just the effect of Zn addition that occurs
 automatically via  the fertilizer loading, even if the pH were kept constant by calcite addition.\\ 
(ii) Even when the available Cd concentration is augmented by various means, this may not be reflected
to the same extent in the plant because the Zn ions will also compete with Cd ions in the rizosphere.
Furthermore, the plant will take up both Cd and Zn ions, and the high Zn component will also be
reflected in the chemical content of the  plant. For instance, taking the rice plant {\it Oriza Satavia},
 a strong phyto-accumulator of Cd  as an example, we show in Table~\ref{rice-tab} a typical 1:1000 Cd/Zn
 ratio in both CKDu-endemic
regions and CKDu-free regions. While the Cd to Zn ratio in the soil may be typically only 1:10 to 1:100,
 the phyto-accumulation of Zn may be much stronger than that of Cd, further increasing
 the plant Zn content
compared to Cd. It is believed that this high  intake
of Zn (and also Se) suppresses the Cd
 intake  in the gut, and may account for the
physiological counteraction of Zn in the diet~\cite{ARLS-review}; and indeed such information
has been available in the literature for perhaps over four  decades~\cite{Jacobs78}.
\begin{table}[t]
\caption{ Concentrations of Cd, Zn and Se present in rice grown in the  endemic `Dry Zone'(DZ)
of Sri lanka (where a form of  chronic kidney disease is found),
and in the `wet zone' (WZ) which is free of the disease. Median amounts have been used where
 possible using the data from~\cite{Diyaba-Rice-2016},
 and \cite{Meharg2013}. The data are for the grain, while the straw  usually has
 2-3 times more Cd and Zn content (n.b. Zn in mg/kg). }
\center
\begin{tabular}{l l l l l l }
\hline\\Rice      &  unit      &     DZ  & DZ$^a$     &     WZ  \\
Cd                & $\mu$g/kg  &     52  & 41.2       &     79  \\
Se                & $\mu$g/kg  &     26  &  -         &     19  \\
Zn                & mg/kg      &     14  & 22.3       &     16  \\
\hline  
\end{tabular}
$\,$

$^a$ mean values, CKDu-endemic area in the DZ, from \cite{Levine16}.
\label{rice-tab}
\end{table}
\subsection{\bf Magnesium and Fluoride mediated enhancement of available Cd in the soil.}
\label{fluo.sec}
~\cite{McLaughlinTil94} drew attention to the impact of salinity and chloride ions on the available
Cd concentration in P-fertilized soils, and proposed that Cd$^{2+}$+Cl${^-}$ complex formation in the
soil solution has to be taken into account as a function of the chloride concentration in the soil
solution. \cite{Smolders01Cd} reported similar results and a linear trend between crop Cd and
 soil Cd.  Similarly, ~\cite{Loga2008} drew attention to the importance of fluoride added to soils
via  P-fertilizer loading, where they considered mainly fluoride toxicity.

 However at the time
 the impact of fluoride ions on the Cd balance, or possible synergies of fluoride, magnesium
 and Cd were not suspected. Recently such
synergies among F, Mg and Cd have been proposed to cause enhanced  nephrotoxicity~\cite{SynergyBandara2017,
CDW17Multiple} via naturally occurring fluoride and hard water in  dug wells rather than from agricultural
inputs. Unfortunately, it is not easy in field trials to control or recognize the role played by 
many  variables  like
fluoride, chloride, and Cd levels etc. Furthermore, glass-house experiments  do not simulate  the
multiple interactions present in actual soils~\cite{CHANEY12}. Of course, results
of simplified experiments can be used in principle to construct the synergies and buffering actions that
come in to play, but in practice this is full of pitfalls.

Most of the multiple ionic interactions occur in the {\it aqueous phase} of the soil solution and hence they
can in fact be treated rather rigorously using methods of electrochemistry and thermodynamics. 
~\cite{Manoharan07} have discussed the complex formation between Al$^{3+}$ ions and fluoride as a function
of soil pH. However, possible  interactions of the fluoride with Cd ions  were not discussed.

In ~\cite{CDW17Multiple} we show by calculations of the change in Gibbs free energy
 that Cd forms a  complex CdF$^+$ which is more stable than CaCl$^+$. Thus the increased presence
of F$^-$ ions in the soil solution will  bring pre-existing exchangeably soil-bound Cd into
soil solution by forming CdF$^+$ ions. This effect can contribute to an observed Cd
enhancement associated with fertilizer addition, as in Fig. 1(a).
 However, while Mg, or Al,
taken individually with fluoride may show complex formation, a mixture of many ions tends to have
 a buffering action on each other, and the effects of multiple ions become less marked. This was
 found to be the case not only from calculations of ionc Gibbs free energies, but also from
 studies of  nephrotoxicity using laboratory mice~\cite{SynergyBandara2017}.

Another aspect of complex formation that we do not discuss in detail in this study is the effect
 of herbicides like glyphosate applied to crops. This leads to a presence of glyphosate and its
breakdown products in the top soil in the short term. They have a salutatory effect on 
heavy-metal content in forming insoluble complexes with, e.g., cadmuim, lead etc., and making
 them not available to plants and soil organisms like earthworms, e.g., by diminishing the
 bio-available soil Cd and making earth worms thrive~(Chui-Fan Zhou et al 2014).
\section{Cadmium content in crops like rice ({\it Oriza sativa})}
\label{rice.sec}
Fig.1 shows the close correlation of the Cd content in soil and in the plant. Although the
rate of uptake of Cd from the soil solution during the growth of a plant depends on  the growth
stage, sunlight, water availability etc., it is possible to make a simple estimate of the final
concentration of Cd, e.g. in paddy and in the water in which it is grown, using a number of
simplifying assumptions. We present  two simple but fairly robust models for the Cd uptake
by a grass or a rice-like plant.

\subsection{Model based on water intake}
We  begin by applying a simplified version of the more detailed analysis (given below)
 to compare the
 predicted Cd uptake-values with the Cd data given in
Table \ref{rice-tab} and reported for Sri Lanka
by~\cite{Diyaba-Rice-2016}.  A
 90-day irrigated  rice crop in the rice-growing north-central province of Sri Lanka
 takes up about 500 mm water~\cite{PriCom17}, i.e., $ 5 \times 10^6$ litres of water
per hectare as an upper bound. Typical values of Cd concentrations are  0.24 $\mu$g/L 
 in canal water~\cite{WHO2}, or 0.11 $\mu$ g of Cd per kg of soil as reported by \cite{Levine16}.
Hence we may take a range of values from 0.5 g - 1.2 g of Cd uptake by paddy per hectare
 per season.

An independent calculation which brings us to consistence with the above numbers is
obtained by looking at the output of paddy (with husk), rice (without husk), straw and stubble,
produced per hectare. Using statistical information for the four years 2011-2015~\cite{CensusStat16}
we have:\\

\begin{tabular}{ll}
Number of hectares averaged over  &= 0.69$\times 10^6$ \\
Average yield of rice,  m.tonnes  &= 3.90 \\
Average yield of husk,  m.tonnes  &= 0.43 \\
Yield of straw+stubble, m.tonnes  &= 4.96 \\
\end{tabular}$\,$\\

Using the above figures and the Cd content $C_{cd}$ of 52 $\mu$g/kg in  the grain,
 2.5$C_{cd}$  and 3$C_{cd}$  for the Cd content in the husk and straw respectively, the
total Cd absorbed from the water works out to an upper bound of about 1.03 g/ha,
 which can be compared to the upperbound  of 1.2 g/ha  Cd estimated from the water intake.

Thus we have consistence  with the experimental data given in  Table \ref{rice-tab}. However,
a more detailed discussion is useful.

Minerals enter the plants through water intake as well as via aerial deposition. Here we ignore the
aerial delivery which may be important in industrial neighbourhoods. The water supply needed
through out the plant's life is used up partly in evaporation, and partly by uptake into the plant.
If the daily water supply is stated as a height $h_w$ (e.g, 0.10 m), the water volume $V_w$ per
hectare is 10$^4\times h_w$ m$^3$ per day. Of this, a fraction $f_e$ is lost by evaporation and the
uptake by the plant is $10^4h_w(1-f_e)$. We define the uptake factor  $f_u=(1-f_e)$.
 At planting and at the initial stages $f_e$ is significant
and may be as high as 50-60\% of that of the grown plant, while most of the water is taken up by the
 plant during its mid-season  growth when the crop is fully developed and in the flowering
 and grain-setting stage. In `dry-harvested' crops like maize, sunflower or paddy, the end-season
 water needs are minimal. Thus $h_w(t)f_u(t)$ are functions of the growth time $t$, which extents
from $t=0$ at planting to $t=T$ at harvesting. Let the Cd concentration in the water
near the roots at the time $t$ be $C_w(t)$. The soil-to-plant transfer coefficient is $f_{sp}(t)$.
 Thus the total mass of Cd (or any other ion) absorbed is
\begin{equation}
M_{\rm Cd}=\int_0^T dt 10^4\times C_w(t)h_w(t)f_u(t)f_{sp}(t).
\end{equation}
If $C_w, f_u, f_{sp}$ are replaced by their average values during growth, and treated
as constants, then we may write the total Cd absorbed by one hectare of crop
during its growth season $T$ as
\begin{equation}
M_{\rm Cd}=10^4\times C_wh_wf_eT =C_wf_eV_w.
\end{equation}
Here $V_w= V_w$ is the total water input during the season. For a 90-day crop requiring
an average of 5mm-7mm per day of water, this amounts to 450-600 mm of water per hectare for
the whole growth period. Using the average values $f_{sp}\simeq 1$,\, $f_e\simeq 0.2,h_w = $7mm
$C_w=0.24 \mu$ g/litre, we can estimate the Cd uptake by one hectare of a rice
plantation during a putative 90-day growth season. Assuming this to yield 4 metric tons of
rice grain, and assuming a distribution of 2:1 or possibly  2.5:1 of Cd between the straw and grain,
 the calculated concentration in the rice grain (30-80$\mu$g/kg) are completely consistent
with the values given in Table {rice-tab}.

Similar calculations can be done for other ions like  Zn or F.  Zn is found in large  excess
 over Cd according to Table \ref{rice-tab}.  Such calculations show  that the measured
concentrations of ions in crops (e.g., as given in Table \ref{rice-tab}) are in {\it grosso modo}
agreement with the concentrations of ions measured in the soil solution, establishing their consistence.

\subsection{Model based on harvest volume}
The rice plant absorbs water and Cd from the ground and grows from a negligible volume
$v_0$ to its final large volume $V_F$ during its life time. The water absorbed
is in fact proportional to this increase in volume $V_F-v_0$.  Let
the volume {\it change } at any moment of its growth  be $dV$. Let the concentration
 of Cd in the neighbourhood of the roots be denoted by $C_w$ at the moment when the plant
has a volume $V$.

Then the amount of Cd absorbed
by the plant in changing its volume by $dV$ is $C_wdV$. There is also a
transfer coefficient $f_{sp}$ connecting the Cd concentration in the soil  and  the Cd
concentration in the plant. As seen from Fig.1 this factor $f_{sp}$ may be taken to be
of the order of  unity
in typical  cases. Hence the total mass of  Cd $M_{\rm Cd}$ absorbed by the plant is:

\begin{equation}
M_{\rm Cd}=\int_{v_0} ^{V_F} f_{sp}.C_w.dV
\end{equation}
If we assume that $f_{sp}$ and $C_w$ can be replaced by their average values during the
lifetime of the plant, we can take them out of the integral sign and write:
\begin{equation}
M_d=C_w.f_{sp}.(V_F-v_0)
\end{equation}
So, neglecting $v_0$, setting $f_{sp}=1$ the Cd absorbed by the plant during its life is
$M_{\rm Cd}=C_wV_F$.
The final volume $V_F$ used here  is the {\it wet volume at  harvest} and not the dry volume.
We consider a crop grown on a hectare of land. Let the average height of a plant to be $h_p$,
while the packing fraction is taken to be $f_p$. Then the volume of plant matter, and also
the weight $ W_F$  of the total wet growth  are  given by
\begin{equation}
 V_F= (1\, \mbox{hectare}) \times (h_p f_p); \;\;\; W_F=V_F\rho.
\end{equation}
In the above, $\rho$ is the density, and may be taken to be close to that of water (i.e., 1 kg per
litre for order-of magnitude calculations).
The packing fraction $f_p$ allows for the fact that there is space among plants unoccupied
by them. In the case of paddy, we may assume that $h_p$ at harvest is 0.5 to 0.75 m, while
the packing fraction $f_p$ may be 0.75 to 0.95  in the full grown condition at harvest time.
The above analysis assumes that the water supply to the soil solution remains more or
less unchanged at saturation level during growth. In  dry-zone cultivation, the soil
water may be cutoff at later stages of growth but such correction effects are indirectly
included in the final plant height and hence on the average high $h_p$  used in the
model. Thus, given experimental values for the quantities needed in the last equation. one
may compare field data with theoretical expectations.

\section{Toxicity effects of Cd in the presence of other ions.}
\label{toxicity.sec}
The neglect of competitive ionic effects seen in many reported experiments is also seen in the
 dietary specifications on  Cd intake. Thus, as already stated, 
 Se, Zn, Mg, and Fe  in the diet have an antagonistic action on  Cd
 toxicity~\cite{ARLS-review,BrzoskaCdZn2001,MatovicCdZnMg2011, CHANEY12} but this is not included
or even alluded to  in specifying the recommended tolerable monthly intake limits (TMIL)
 on  Cd in the
diet as indicated in, say, CODEX alimentarius stipulations.
Of course, local authorities have the freedom to re-interpret the TMIL to mean that if the Zn inputs
 are over-overwhelmingly large, then the Cd inputs may be  ignored. This usually happens mainly on the
 strength  of tradition rather than on the basis of science.
  Sunflower kernels and other foods like shellfish are high in Cd and yet show no adverse effects when
consumed~\cite{Sirot08Shell}. Farming communities in the UK in regions with high
Cd in the soil consume diets rich in potatoes and cereals without any adverse effects \cite{CHANEY12}.
 Similarly, the lack of 
chronic Cd toxicity in many  communities, where  rice containing  Cd in amounts exceeding the
TMILs has been consumed for generations, can be explained by the protective action of adequate amounts
of ions like Zn, Se or Fe in the diet (see  Dharma-wardana, 2017, and Sec. 5.4 of Chaney, 2012).
 Conversely, when Cd-toxicity from crop products
occurs, it is mostly likely that the diet is grossly deficient in protective micronutrients like Zn or Se. 
Table~\ref{rice-tab} shows that Sri Lankan Chronic Kidney disease is uncorrelated with Cd in rice.
Hence other explanations have been conidered \cite{SynergyBandara2017, Thammiti17}.               
The Codex Alimentarius~\cite{Codex18}  uses a single- variable step-function model for stipulating
a chronic toxicity-onset amount $m_{\rm Cd}$ per kg of body weight  per day, week or month, as is appropriate.
No synergies or counter-effects of other contaminants are included in the specification.
If for example the daily inputs of Cd, and Zn, Fe ... are $I_{\rm Cd}, I_j$, $j$ = Zn, Fe, only the
amounts scaled by their
bioavailable fractions $f^a_j$ are of importance. Many studies, e.g., ~\cite{Premarathne2006},
 Smolders et al \cite{Smolders13}, 
 show that $f^a_j$ if the order of 30-50\% for common vegetables, rice, etc.,
i.e., $f^a_j\sim0.4$.
Furthermore, each ion has an uptake factor $f^u_j$ for intestinal absorption.
Only about 2.5-6\% of the bioavailable Cd  is absorbed in the intestines, with $f^u_{\rm Cd}\sim0.05$
 \cite{ATSDR2008Cd}.
 According to \cite{kim07Cd} Cd absorption in the gut involves a ferrous transporter,
which also takes up Zn, while Zn has other transporters associated with its uptake, and hence
the details are unclear \cite{Iyengar09}.
 Furthermore, iron deficiencies can cause higher Cd absorption. Both ferrous ions and
Zn ions are believed to be more actively taken up by this transporter, but even if we assume that the uptake
 factors $f^u$ are the same for the three elements, the Cd uptake will be reduced to a third or less if ferrous and Zn ions
are present in equal amounts to compete with Cd. That is, using the simplest picture  (i.e., without including synergies),
it is only if the potential amount of Cd available in the gut  for uptake
exceed the total amount of its competitor ions that there would be absorption.
That is, it is reasonable to conclude that the  condition
\begin{equation}
I_{\rm Cd}f^a_{\rm Cd}f^u_{\rm Cd} > 
  \Sigma_jI_jf^a_jf^u_j
\end{equation}
has to be satisfied for any significant Cd absorption by the gut to set in.

\section{Conclusion}
We have reviewed the widely held hypothesis that  ``soil-Cd concentrations get  enhanced by the
use of P-fertilizer at rates which are likely to create  dangerous conditions for human health in
a few  decades'',
 and conclude the following.
(a) This strong concern is not justified at current levels of fertilizer usage where the doubling
of the Cd cointent in soils would take centuries and not decades. 
(b) The causes of  increased bioavailable soil Cd on fertilizer addition are most probably
 found in other factors  that cause the release of {\it pre-existing}  Cd found in the soil.

 These conclusions
follow since the incremental change in the bio-available soil Cd concentration on
addition of P-fertilizer is in fractions of mg/kg of soil per year, while ambient
 soil Cd levels are millions of times larger.
 The factors that cause the increase in bio-available soil Cd are most likely
 to be the following. (i) Change in soil pH due to fertilizer action and associated  action
 of micro-organisms,
(ii) The effect of ionic forms of  Zn, Mg, F, Cl, Ca, etc., on the ionic equilibria of the soil solution,
  given that   such ionic forms are found in P-fertilizers, ag-lime  and such agrochemicals.
(iii) Competitive effects on clay adsorption sites, humic acid moieties, and in the rizosphere, 
(iv) Proliferation of the root system and its activity  under fertilizer additon,
 leading to increased dissolved
Cd in the soil (and in the plant).
(v)  Ionicity effects on  organic matter and other effects that we have not discussed in this study.

 We have also pointed out that the neglect of ion synergies (e.g., Zn in suppressing Cd toxicity) in
 specifying tolerable maximum weekly intake values can lead to paradoxical situations where healthy
 communities have been found to be  consuming diets that would appear to be dangerous to health
 if  the CODEX  alimentarius stipulations are applied naively.

 Furthermore, we conclude that attempting to  control
 the enhancement of bio-available Cd in soils caused by P-fertilizer loading may require
 controlling their fluoride, magnesium, and Zn  content rather than the Cd content.
Reducing potato diets in favour of wheat and rice diets would cut  down fertilizer inputs
perpahs by a factor of four.  In addition, the push by the European Food Safety Agency
  (as well as similar  organizations)  to reduce the Cd content in crops
 by continued  lowering of the allowed Cd levels in  P-fertilizers would turn out to be an
  expensive  and futile exercise. 

{\it Acknowledgement} -- The author thanks Dr. Sarath Amarasiri for his comments and
 drawing attention to some references. 
 

\end{document}